\def\year{2021}\relax
\documentclass[letterpaper]{article} 
\usepackage{pdfpages}
\usepackage{aaai21}  
\usepackage{times}  
\usepackage{helvet} 
\usepackage{courier}  
\usepackage[hyphens]{url}
\usepackage{hyperref}
\usepackage{graphicx} 
\usepackage{subfiles}
\usepackage{biblatex}
\usepackage{caption}
\title{DEAP Cache: Deep Eviction Admission and Prefetching for Cache}
\author{
    Ayush Mangal,\textsuperscript{\rm 1}\thanks{All authors contributed equally, the names are listed in alphabetical order}  Jitesh Jain,\textsuperscript{\rm 1}\footnotemark[1] Keerat Kaur Guliani,\textsuperscript{\rm 1}\footnotemark[1] Omkar Bhalerao\textsuperscript{\rm 1}\footnotemark[1]
    \\
}
\affiliations{
    \textsuperscript{\rm 1} Vision \& Language Group, IIT Roorkee\\
    amangal@cs.iitr.ac.in,
    jitesh\_j@cs.iitr.ac.in, kguliani@ce.iitr.ac.in,
    obhalerao@me.iitr.ac.in
    
}

\begin{document}

\maketitle

\begin{refsection}[library.bib]
\begin{abstract}
Recent approaches for learning policies to improve caching, target just \textit{one} out of the \textit{prefetching}, \textit{admission} and \textit{eviction} processes. In contrast, we propose an end to end pipeline to learn \textit{all} three policies using machine learning. We also take inspiration from the success of pretraining on large corpora to learn specialized embeddings for the task. We model prefetching as a sequence prediction task based on past misses. Following previous works suggesting that frequency and recency are the two orthogonal fundamental attributes for caching, we use an online reinforcement learning technique to learn the optimal policy distribution between two orthogonal eviction strategies based on them. While previous approaches used the past as an indicator of the future, we instead explicitly model the future frequency and recency in a multi-task fashion with prefetching, leveraging the abilities of deep networks to capture futuristic trends and use them for learning eviction and admission. We also model the distribution of the data in an online fashion using Kernel Density Estimation in our approach, to deal with the problem of caching non-stationary data. We present our approach as a "proof of concept" of learning \textit{all} three components of cache strategies using machine learning and leave improving practical deployment for future work. 
\end{abstract}.

\section{Introduction}
Caches having low latency have limited space, which must be utilized effectively. Since the problem of accessing such data from the main memory is predictive in nature, various efforts have previously been directed to applying machine learning techniques to the task of cache optimisation.\\
\cite{Hashemi2018} modelled the task of prefetching as a sequence prediction problem based on past misses, which we adopt as well. \cite{Vietri2018a} demonstrated \textit{frequency} and \textit{recency} as two orthogonal attributes for cache eviction decisions and learned an optimal policy distribution between two approaches based on \textit{past} estimates namely LRU and LFU. \cite{liu2020imitation} used an imitation learning-based approach for cache replacement, wherein they used a byte-level representation to deal with the exponential size of address vocabulary. 
They also observed that learning \textbf{both} prefetching and replacement had not been appropriately addressed in any previous work.
Improving on these approaches, we address the main contribution of this paper as :
\begin{itemize}
    \item We propose a machine learning method to learn all \textbf{three} components of caching strategies, i.e., prefetching, admission and replacement.
    \item We enhance the byte level representations using recent advances in natural language processing.
    \item We tackle the problem of non-stationary data by modelling the data distribution explicitly using Kernel Density Estimation (KDE).
    \item We explicitly model the future estimates of two orthogonal attributes, namely frequency and recency, for learning the optimal replacement and admission policies, instead of using the past as an indicator of the future.
\end{itemize}

\begin{figure*}
    \centering
    \includegraphics[width=\linewidth,scale=0.50]{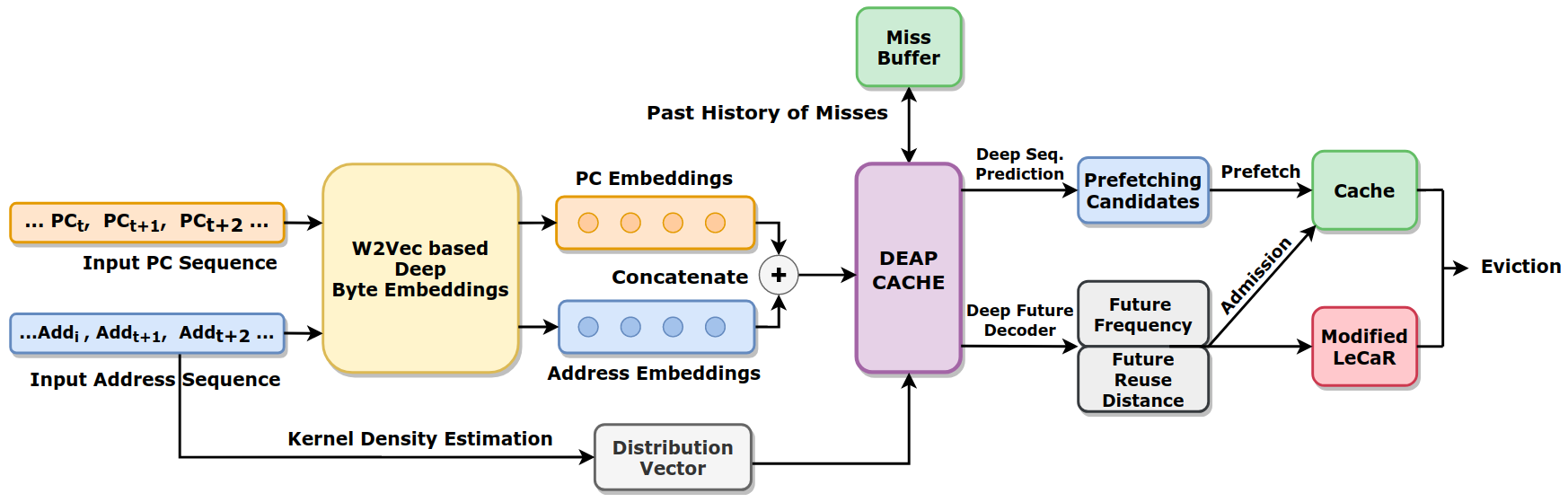}
    \caption{Schematic diagram of our approach. We feed specialised embeddings extracted from input address sequence into our DEAP Cache model to make admission, prefetching and eviction decisions.}
    \label{fig:schematic}
\end{figure*}

\section{Methodology}
\subsection{Training Mode (Offline)}
\textbf{\textit{Pretrained Byte Embeddings using Word2Vec:}} Following \cite{liu2020imitation}, we use byte-level embeddings of the two features used to represent cache misses: the missed address and the corresponding Program Counter (PC). However, we take it one step further, deriving from recent advances in pretraining on text corpora to train Word2Vec \cite{mikolov2013distributed} based specialized byte embeddings.\\
\textbf{\textit{Sequence Modelling for Prefetching Candidates:}} The sequence of the obtained "miss" embeddings is passed through a Long Short-Term Memory (LSTM) network to get a probability-wise prediction of the expected (subsequent) cache misses to be prefetched.\\
\textbf{\textit{Sequence Distribution Estimation:}}
We deal with the problem of non-stationarity of the data to be cached, by explicitly modelling the current distribution of the sequence using a non-parametric method called Kernel Density Estimation, and feed the resultant distribution vector into the pipeline.\\
\textbf{\textit{Multitasking Frequency and Reuse Distance Prediction with Prefetching:}}
Unlike previous works, we model \textbf{future} frequency and reuse distance (timesteps till next occurence) by applying a learnable decoder to the embedding of the address in question, along with the current estimate of the distribution, in a multi-task fashion with prefetching prediction.
\subsection{Testing Mode (Online)}
 \textit{\textbf{Admission Policy:}} We use the decoder mentioned in the previous section to predict an estimate of the future recency/frequency of the address and then use a threshold to decide whether to admit the address or not.\\
\textit{\textbf{Prefetching Policy:}} We maintain an online buffer of the past $k$ misses and pass samples from it in every $T$ timesteps to the LSTM model to get candidates for prefetching.\\
\textit{\textbf{Eviction Policy:}} We modify the LeCaR approach of \cite{Vietri2018a} and use the concept of \textit{regret minimization} to learn the optimal probability distribution between two eviction policies, one based on \textit{future} recency and other on \textit{future} frequency. Note that \cite{Vietri2018a} instead used LRU and LFU that modelled the future based on \textit{past} metrics. We refer the reader to the supplementary for a detailed description of our approach.

\section{Experiments and Results}
To test the validity of our approach, we considered five baseline approaches for evaluation: LRU, LFU, FIFO, LIFO, Oracle. We used a free publicly available dataset\footnote{We derived our dataset from the dataset found  \href{https://cseweb.ucsd.edu/classes/sp14/cse240A-a/project1.html}{here}} due to financial constraints as students. As can be seen in \autoref{tab:my-table}, our approach supersedes the \textit{Mean Hit Rate} obtained by all previous classical approaches and comes the closest in performance to the optimal figure obtained from BELADY's algorithm (Oracle) \cite{5388441}, thus demonstrating the validity of our approach. We open-sourced the code\footnote{The codebase and dataset used can be found \href{https://github.com/vlgiitr/deep_cache_replacement}{here}}
 and provide a detailed account for reproducibility in the supplementary.
\begin{table}
\centering
\begin{tabular}{l|l}
\hline
\textbf{Method} & \textbf{Mean Hit Rate} \\ \hline
LRU    & 0.42     \\
LFU    & 0.43     \\
FIFO   & 0.36     \\
LIFO   & 0.03    \\
BELADY (Oracle) & 0.54     \\
\textbf{Ours}   & \textbf{0.48}     \\ \hline
\end{tabular}
\caption{Different approaches and their mean hit rates}
\label{tab:my-table}
\end{table}

\section{Conclusion \& Future Work}
In this work, we proposed an end to end pipeline for learning \textit{all} the three components of caching strategies using machine learning and demonstrated the superiority of our approach over classical baselines. Improving our approach's practical deployment and evaluating on large-scale real-time benchmarks is an interesting future direction.
\printbibliography
\end{refsection}
\clearpage
\begin{refsection}[suppl.bib]

\setcounter{secnumdepth}{0}
\section{Supplementary}
 \vspace{0.5cm}
\section{Implementation Details}
We will now describe the implementation of our approach in detail. We will first describe the training phase, and then the online testing simulation.
\section{Training}
The training phase happens in an offline fashion, where we sample sequences from our dataset and carry out simulations to train our model. It is mainly concerned with:
\begin{itemize}
\item Pretraining of specialised byte embeddings for the domain using word2vec based approach.
\item Training of an LSTM model for predicting future miss addresses for prefetching based on the sequence of past misses.
\item Estimating the current distribution of the address sequence in a quick online fashion.
\item Training a decoder to predict the \textit{future} reuse distance ( timesteps till next occurence) and frequency estimates in a multi-task fashion with prefetching candidate preditction.
\end{itemize}
 
\subsection{Pretrained Byte Embeddings using Word2Vec}
\label{subsec:byteembeddings}
Following \cite{liu2020an}, to deal with the exponential size of the vocabulary of all possible unique addresses (for example, $2^{32}$ possible unique addresses in a 32-bit address system), we model the addresses and PCs using byte-level embeddings. Going one step ahead, we take inspiration from recent advances in pretraining word embeddings and use a variant of word2vec \cite{mikolov2013distributed} to get pretrained byte-level embeddings using a large corpora of addresses and PCs generated via multiple program simulations. While using these byte-level embeddings in our model, we apply a learnable multi-layer perceptron (MLP) above the concatenated byte embeddings, which combines the byte level embeddings to learn an "address level" embedding in a manner that is specially optimized for the downstream tasks. For example, for a 4 byte address $A$, represented in bytes as $(B1,B2,B3,B4)$ we get byte embeddings $b_{j}$ and address-level embeddings $a$ as:
\begin{center}
$b_{j} = W_{j}^{T}b_{j} \qquad j\in {1\cdots4}$ \\
$a = f(b1 \oplus b2  \oplus b3  \oplus b4)$
\end{center}
Where $W_{j}$ are the corresponding embedding matrices trained using w2vec, $\oplus$ represents the concatenation of the byte embeddings, and $f()$ represents the MLP used to convert the byte level embeddings to an address level embedding.

\subsection{Sequence Modelling for Prefetching Candidates}
\label{prefecthing}
Following \cite{Hashemi2018}, we model the problem of prefetching as a sequence prediction task using the past history of cache miss addresses $A^{m}$ and their corresponding program counters $P^{m}$. Given an input sequence of miss addresses $[A^{m}_1, A^{m}_2 \dots, A^{m}_{k}]$ and corresponding PC addresses $[P^{m}_1, P^{m}_2 \dots, P^{m}_{k}]$, the aim is to predict a set of $n$ most probable prefetching candidate addresses $[C_1, C_2 \dots, C_{n}]$.\\
We generate the miss address embeddings $[a_1, a_2 \dots, a_{k}]$ and PC embeddings $[p_1, p_2 \dots, p_{k}]$ from the input sequence using the method described in this \autoref{subsec:byteembeddings}. We then concatenate the missed addresses' embeddings and missed PCs' embeddings to get the input embeddings 
$[e_1, e_2 \dots, e_{k}]$.

\begin{center}
    $e_{i} = a_{i}\oplus p_{i}$
\end{center}
The input embeddings are then fed into a Long Short Term Memory (LSTM) model which captures both the short and long term dependencies across the sequence to predict future miss addresses \cite{hochreiter1997long} better.

\begin{center}
        $\overrightarrow{h_{i}}=LSTM^{(f)}(e_i, \overrightarrow{h}_{i-1})$
\end{center}

The hidden state from the last step $h_{k}$ is then fed into a dense layer $g_{j} (j\in {1\cdots4})$ with a softmax applied on top, to get the probability distribution over the future miss address' predictions. Note that to deal with the exponential size of the addresses, we predict each byte $\hat{b}_{j}$ separately.

\begin{center}
    $\hat{b}_{j} = softmax(g_{j}(h_{k})) \qquad j\in {1\cdots4}$
\end{center}

Where $g_{j}$ is the dense layer to predict the byte $b_{j}$. We use the Cross-Entropy loss to train our prefetching candidate prediction model, which is given by:

\begin{center}
    $\mathcal{L}_{prefetching} =  -(\frac{1}{n})\sum_{i=1}^{n}\sum_{j=0}^{4}\sum_{c=0}^{255}b_{i,j,c}\log(\hat{b}_{i,j,c})$
\end{center}

\subsection{Sequence Distribution Estimation}
\label{distribution}
The input address sequence distribution is bound to be non-stationary in the real world online setting where caches are usually deployed. To deal with this, we propose explicitly modelling the current distribution of the sequence and providing an inductive bias to help our model deal with the problem of moving address distribution. To make the approach quick and online, we refrain from using deep learning methods to estimate the distribution and instead used a classical non-parametric way of probability density estimation, i.e., \textit{Multivariate Gaussian Kernel Density estimation} \cite{kristan2011multivariate}. Given an input sequence of embeddings $[e_1, e_2 \dots, e_{k}]$, we obtain the distribution vector $d_{i}$ corresponding to the sequence by applying Kernel Density Estimation (KDE) on it :

\begin{center}
    $d_{i} = KDE(e_1, e_2 \dots, e_{k})$
\end{center}

\subsection{Multitasking Frequency and Reuse Distance Prediction Decoder with Prefetching Candidate Prediction}
\label{multitask}
Following recent advances in multi-task learning \cite{caruana1997multitask} indicating that \textit{positive transfer} from various related tasks improves performance, we apply multi-task learning to the problem of predicting an estimate of the "future" frequency and reuse distance of an address in question along with the task of prefecthing candidate prediction. Reuse distance is defined as the number of timesteps until the next occurence of the address, and represents a future "recency" of the address. A similar approach was followed in \cite{liu2020an}, but it was only limited to predicting reuse distance. We extend their approach to predicting both an estimate of the future frequency and the reuse distance in a multi-task fashion along with the prefetching predictions. Note that the distribution vector representing an estimate of the current distribution plays a vital role in this prediction, and thus we concatenate the distribution vector $d_{i}$ with the address embedding  $a_{i}$ and then feed it to an MLP based decoder with multiple heads. This gives us an estimate of the future frequency $f_{i}$ and reuse distance  $r_{i}$:

\begin{center}
    $z_{i} = a_{i} \oplus d_{i} $\\
    $f_{i} = F(z_{i}) $\\
    $r_{i} = R(z_{i}) $\\    
\end{center}

Where $F()$ and $G()$ are the frequency and reuse distance prediction  MLP head, respectively. To train this decoder in an end-to-end fashion along with our prefetching module, we would need to use the addresses predicted by the module. However, this would be problematic, since the prefetching module outputs a probability distribution instead of a particular address, and taking an $argmax()$ to get the most probable prediction would introduce non-differentiability ( which is undesirable for an end-to-end pipeline that relies on backpropagating on the prefetching model as well). To bypass the non-differentiability of argmax, we instead use a temperature-based approach to convert the soft probability into an approximate argmax, which is then utilised to calculate the address embeddings during training. However, note that this is not required during testing since we already have access to  eviction/addmission address candidates for which we need to make frequency-recency estimates.

We use the mean squared error (MSE) loss between the predicted frequency/reuse distance and the ground truth to train the decoder, which is given by:\\
\begin{center}
    $\mathcal{L}_{frequency} = (\frac{1}{n})\sum_{i=1}^{n}(\hat{f}_{i} - f_{i})^{2}$\\
    $\mathcal{L}_{recency} = (\frac{1}{n})\sum_{i=1}^{n}(\hat{r}_{i} - r_{i})^{2}$
\end{center}

Where $\hat{f}_{i}, {f}_{i}$ and $\hat{r}_{i}, {r}_{i} $ are the predicted and ground truth frequency and reuse distance respectively. The total loss for training is given by a weighted average of the three losses:\\
\begin{center}
     $\mathcal{L}_{total} = w_{0}\mathcal{L}_{prefetching} + w_{1}\mathcal{L}_{frequency} + w_{2}\mathcal{L}_{recency}  $
\end{center}

\section{Testing Simulation Pipleine}
Consider an input sequence of addresses $[A_1, A_2 \dots, A_{T}]$ and corresponding PC addresses $[P_1, P_2 \dots, P_{T}]$. We need a caching strategy to cache them efficiently in a cache $C$ of size $s$, to maximise the number of hits (hit-rate). Any caching strategy would consist of three main components :
\begin{itemize}
\item \textbf{Admission Policy} - Deciding whether to cache an address after it causes a miss.
\item \textbf{Prefetching Policy} - Predicting the addresses which \textit{will} lead to misses in the future and "prefetching them".
\item \textbf{Eviction Policy} - Deciding which address to evict to make space for new addresses to be admitted into the cache.
\end{itemize}

\begin{table*}[!ht]
\centering
\begin{tabular}{l |l |l}
\hline
\textbf{Hyperparameter}              & \textbf{Search Space}                                      & \textbf{Optimal Choice} \\ \hline
Number of Epochs           & uniform-integer{[}1, 20{]}                                 & 20                       \\
Training Batch Size                 & choice{[}32, 64, 128, 256, 512{]}                             & 256                     \\
Optimizer                  & choice{[}Adam, SGD{]}                                       & Adam                    \\
Learning Rate              & loguniform-float{[}1e-5, 1e-1{]}                          & 1e-3                    \\
Training Temperature     & choice{[}1e-3, 1e-2{]}                                    & 1e-3                    \\
LSTM Hidden Cell Size      & uniform-integer{[}20, 40{]}                                & 40                      \\
Decoder Hidden Size  & 10                                                         & 10                      \\
Prefetching Input Sequence Length          & choice{[}20, 30{]}                                          & 30                      \\
Address Embedding Size              & uniform-integer{[}5, 25{]}                                 & 20            \\
Weight for Cross Entropy Loss            & 0.33                                                       & 0.33                    \\
Weight for Frequency MSE-Loss      & 0.33                                                       & 0.33                    \\
Weight for Reuse Distance MSE-Loss  & 0.33                                                       & 0.33  \\
Word2Vec Number of Epochs               & uniform-integer{[}20, 500{]}                                & 120                     \\
Word2Vec Learning Rate                  & loguniform-float{[}1e-5, 1e-2{]}                          & 3e-3                    \\
Word2Vec Weight Decay                   & loguniform-float{[}1e-6, 10{]}                            & 1e-3                    \\
Word2Vec Optimizer                      & choice{[}Adam, SGD{]}                                       & Adam                    \\
Word2Vec Encoder Hidden Layer Size      & uniform-integer{[}50, 200{]}                                & 128                     \\
Word2Vec Byte Embedding Dimension            & uniform-integer{[}5, 25{]}                                 & 20                      \\
Word2Vec Context Size                   & uniform-integer{[}2, 10{]}                                 & 4                       \\
Admission Frequency Threshold ($\alpha$)     & choice{[}50, 300, 500, 1000,3000{]}                                          & 3000                      \\
Admission Reuse Distance Threshold ($\beta$) & choice{[}500, 3000, 5000, 7000,8000{]}                                       & 7000                      \\
Miss Buffer Size     & choice{[}30, 50, 70, 100{]}                                          & 50                      \\
Test Simulation Prefetching Interval & choice{[}10, 20, 30, 50{]}                                       & 30                      \\
Cache Size           & choice{[}32, 64{]}                                          & 32                      \\
Test Simulation Batch Size           & choice{[}5000, 10000{]}                                     & 10000                   \\ \hline
\end{tabular}
\caption{Hyperparameter search space for our model}
\label{hyper-parameters}
\end{table*}

We now describe all three components of our strategy:

\subsection{Admission Policy} 
Given an input PC ($P_{t}$) and Address ($A_{t}$), we first embed them into an embedding $e_{t}$ as described in \textit{this \autoref{subsec:byteembeddings}}. We then predict its estimated frequency $f_{t}$ and reuse distance $r_{t}$ using the decoder trained as described in \textit{this \autoref{multitask}}. We then admit only those addresses which have their frequency above a threshold $\alpha$ (to admit more frequent addresses) or reuse distance below a threshold $\beta$ (to admit addresses which will occur soon because of shorter reuse distance). Thus the admission decision $y_{t}$ is given by:

\begin{center}
   $ y_{t}= \biggl\{\begin{array}{@{}l@{}l@{}}
    
    1, f_{t} > \alpha \, or \, r_{t} < \beta
  ,\\
   0, otherwise
  \end{array}$
\end{center}

\subsection{Prefetching Policy}
We maintain an offline buffer consisting of the past $k$ miss addresses and their corresponding PC addresses. During the simulation, after every $T$ timesteps, we sample both of them from the "miss buffer" and pass them into the LSTM model as described in \textit{this \autoref{prefecthing}}. The model outputs a probability distribution of the candidates to be prefetched. We then prefetch the $n$ most probable candidates by appropriately sampling from this distribution.

\subsection{Online Learning of Eviction using Modified LeCaR}
As mentioned in \cite{vietri2018driving}, there are two fundamental and orthogonal characteristics of elements for caching namely \textit{frequency} and \textit{recency}. Correspondingly, there are two fundamental eviction strategies LRU ( Least Recently Used) and LFU ( Least Frequently Used). Hence \cite{vietri2018driving} proposed an online method of \textbf{Le}arning
\textbf{Ca}che \textbf{R}eplacement (LeCaR) using the concept of \textit{regret minimisation} \cite{zinkevich2008regret} to learn the optimal probability distribution among the two orthogonal policies of LRU and LFU. 

However, both these eviction strategies, as used in LeCaR, used the past frequency/recency as an estimate of the corresponding future values, which is based on the assumption that the past is a good indicator of the future. We feel this assumption can prove to be questionable. We propose to improve this model by explicitly predicting the future frequency and recency ( via reuse distance) using the decoders $F()$ and $R()$ as described in \textit{this \autoref{multitask}}. We then apply regret minimization to choose the optimal probability distribution between the policies of evicting the address with the least predicted future frequency and the address with the highest predicted reuse distance in the future. We claim that using the predicted future values from our DL model serves as a better estimate of the future than the past. We refer the reader to the original paper \cite{vietri2018driving} for more details about the original approach.

\section{Reproducibility Index}
In this section, we provide a detailed description of our experimentation settings, the baselines used, and dataset details to improve our method's reproducibility. As students, we have limited access to computational resources and commercial hardware benchmarks and hope that making our approach reproducible would allow evaluating and improving our approach's practical deployment. We have open-soruced our codebase and the dataset used at (\url{https://github.com/vlgiitr/deep_cache_replacement}).

\section{Baselines}
We now desribe the baseline against which our model was compared :
\begin{itemize}
\item \textbf{Least Recently Used (LRU):} A fundamental technique based on recency in which we evict the candidate that has been the least recently used.
\item \textbf{Least Frequently Used (LFU):} A fundamental technique based on frequency in which we evict the candidate that has been the least frequently used.
\item \textbf{First in First Out (FIFO):} A heuristic-based technique in which we evict the candidate which entered the cache the earliest.
\item \textbf{Last in First Out (LIFO):} A heuristic-based technique in which we evict the candidate which entered the cache the most recently.
\item \textbf{BELADY (Oracle):} A theoretically optimal oracle \cite{5388441} which has access to the entire future sequence and takes an optimal decision based on it.
\end{itemize}

\subsection{Hyperparameters}
We provide a detailed description of the hyper-parameters search space used to train our approach in \autoref{hyper-parameters} for enhancing reproducibility. We describe the following hyperparameters :
\begin{itemize}
\item \textbf{Number of Epochs:}  The number of epochs for which the training was carried out.
\item \textbf{Training Batch Size:} The batch size for the training.
\item \textbf{Optimizer:}  The optimizer used to improve the model through backpropagation.
\item \textbf{Learning Rate:} The learning rate used to train the model.
\item \textbf{Training Temperature:} The temperature used to perform argmax in a differentiable manner.
\item \textbf{LSTM Hidden Cell Size:}  The dimension size of the hidden state of the LSTM.
\item \textbf{Decoder Hidden Size:} The dimension size of the hidden layer of the decoder used to get the frequency and reuse distance of an address. 
\item \textbf{Prefetching Input Sequence Length :}  The number of past miss addresses and PC address taken in the sequence which is then fed into the LSTM model as input.
\item \textbf{Address Embedding Size:} The dimension size of the encoded address embeddings using an MLP on top of the byte embeddings.
\item \textbf{Weight for Cross Entropy-Loss:} The relative weight of the cross entropy-loss for prefetching candidate prediction in the total loss.
\item \textbf{Weight for Frequency MSE-Loss:} The relative weight of the frequency MSE-loss in the total loss.
\item \textbf{Weight for Reuse Distance MSE-Loss:} The relative weight of the reuse distance MSE-loss in the total loss.
\item \textbf{Word2Vec Number of Epochs:}  The number of epochs for which the word2vec based pretraining was carried out for the byte embeddings.
\item \textbf{Word2Vec Learning Rate:} The learning rate used in Word2vec.
\item \textbf{Word2Vec Weight Decay:} The weight decay used to provide regularization during word2vec.
\item \textbf{Word2Vec Optimizer:}  The optimizer used in word2vec.
\item \textbf{Word2Vec Encoder Hidden Layer Size:} The dimension size of the hidden layer of the encoder used to get the byte embedding. 
\item \textbf{Word2Vec Context Size:}  The number of surrounding bytes used as input while training using word2vec.
\item  \textbf{Admission Frequency Threshold:} The frequency threshold value above which admission occurs.
\item \textbf{Admission Recency Threshold:} The recency threshold value below which admission occurs.
\item \textbf{Miss Buffer Size:} The number of past address misses and program counters used as input during prefetching.
\item \textbf{Test Simulation Prefetching Interval:} The number of timesteps after which we run one instance of prefetching.
\item  \textbf{Cache Size:} The size of the cache used for test simulation.
\item \textbf{Test Simulation Batch Size:} The batch size used for carrying out the test simulation.
\end{itemize}

\subsection{Dataset}
The standard benchmark dataset for training and evaluation of cache simulations is the SPEC CPU2000 benchmark \cite{henning2000spec}, which costs \$2000 and hence was out of the financial scope of the student authors. So, we used a proxy dataset available at \href{https://cseweb.ucsd.edu/classes/sp14/cse240A-a/project1.html}{(https://cseweb.ucsd.edu/classes/sp14/cse240A-a/project1.html)} as a reference for preparing our datasets. We created five files with sequence length ranging from 180000 to 500000 using the preceding URL. Testing the validity of our approach on the SPEC CPU2000 benchmark is an interesting future direction.

\printbibliography
\end{refsection}
\end{document}